\documentclass[
reprint,
superscriptaddress,
bibnotes,
amsmath,
amssymb,
aps,
prl,
]{revtex4-1}
\usepackage{graphicx}
\usepackage{dcolumn}
\usepackage{bm}
\usepackage[final]{changes}
\makeatletter
\@namedef{Changes@AuthorColor}{red}
\colorlet{Changes@Color}{red}
\makeatother
\usepackage[colorlinks,
			linkcolor=blue,
            anchorcolor=blue,
            citecolor=blue,
            urlcolor=blue,
            breaklinks=true]{hyperref}

\newcommand{\eref}[1]{Eq.~(\ref{#1})}
\newcommand{\fref}[1]{Fig.~\ref{#1}}

\newcommand{\ket}[1]{\ensuremath{|#1 \rangle}}
\newcommand{\expv}[1]{\ensuremath{\langle #1 \rangle}}

\newcommand{\e}{\ensuremath{\varepsilon}}
\newcommand{\p}{\ensuremath{\partial}}
\newcommand{\im}{\ensuremath{\mathrm{i}}}
\newcommand{\df}{\ensuremath{\mathrm{d}}}

\newcommand{\calR}{\ensuremath{\mathcal{R}}}
\newcommand{\calT}{\ensuremath{\mathcal{T}}}
\newcommand{\calV}{\ensuremath{\mathcal{V}}}
\newcommand{\calG}{\ensuremath{\mathcal{G}}}

\newcommand{\bR}{\ensuremath{\mathbb{R}}}

\newcommand{\bk}{\ensuremath{\mathbf{k}}}

\newcommand{\br}{\ensuremath{\mathbf{r}}}

\newcommand{\calX}{\ensuremath{\mathcal{X}}}
\newcommand{\calE}{\ensuremath{\mathcal{E}}}
\newcommand{\calF}{\ensuremath{\mathcal{F}}}
\newcommand{\ie}{\emph{i.e.}}
\newcommand{\eg}{\emph{e.g.}}

\begin{document}

\title{Inhomogeneity-Induced Casimir Transport of Nanoparticles}
\author{Fanglin Bao}
\affiliation{Centre for Optical and Electromagnetic Research, Guangdong Provincial Key Laboratory of Optical Information Materials and Technology, South China Academy of Advanced Optoelectronics, South China Normal University, Guangzhou 510006, China}
\author{Kezhang Shi}
\affiliation{Centre for Optical and Electromagnetic Research, JORCEP, Zhejiang University, Hangzhou 310058, China}
\author{Guanjun Cao}
\affiliation{Centre for Optical and Electromagnetic Research, Guangdong Provincial Key Laboratory of Optical Information Materials and Technology, South China Academy of Advanced Optoelectronics, South China Normal University, Guangzhou 510006, China}
\author{Julian S. Evans}
\affiliation{Centre for Optical and Electromagnetic Research, JORCEP, Zhejiang University, Hangzhou 310058, China}
\author{Sailing He}\email{sailing@kth.se}
\affiliation{Centre for Optical and Electromagnetic Research, Guangdong Provincial Key Laboratory of Optical Information Materials and Technology, South China Academy of Advanced Optoelectronics, South China Normal University, Guangzhou 510006, China}
\affiliation{Centre for Optical and Electromagnetic Research, JORCEP, Zhejiang University, Hangzhou 310058, China}
\affiliation{Department of Electromagnetic Engineering, Royal Institute of Technology, 10044 Stockholm, Sweden}
\date{May 9, 2018}

\begin{abstract}
This letter proposes a scheme for transporting nanoparticles immersed in a fluid, relying \deleted{solely }on \replaced{quantum vacuum}{molecular and charge} fluctuations. The mechanism lies in the inhomogeneity-induced lateral Casimir force between a nanoparticle and a gradient metasurface, and the relaxation of the conventional Dzyaloshinski\v{\i}-Lifshitz-Pitaevski\v{\i} constraint\deleted{ due to superhydrophobicity}, which allows quantum levitation for a broader class of material configurations. The velocity for a nanosphere levitated above a grating is calculated and can be up to a few microns per minute. The Born approximation gives general expressions for the Casimir energy \replaced{which reveal size-selective transport. For any given metasurface, a certain particle-metasurface separation exists where the transport velocity peaks, forming a ``Casimir passage".}{ and the lateral force, while exact simulations show the contribution of multipole harmonics to the lateral force is nonmonotonic with respect to the order of harmonics. Additional dependence of the lateral force on the filling factor, which is beyond the Born approximation, is observed. Size dependence of the transport velocity for different nanospheres is analysed.} The sign and strength of the Casimir interactions can be tuned by the shapes of liquid-air menisci, potentially allowing real-time control of an otherwise passive force, and enabling interesting on-off or directional switching of the transport process.
\end{abstract}

\maketitle

\deleted{Nanoparticles experience different significant forces than macroscopic objects. Careful consideration of these forces can lead to new opportunities for developing technologies or explaining fundamental physical, biological or chemical processes. While fluctuations usually lead to disorder and the well-known Brownian motion is spherically symmetric in most cases, the idea of a Brownian motor (with an external auxiliary field) plays a critical role in ATP biochemistry . }Controlling nanoparticles is an essential tool that allows for an improved understanding of nanoscale forces and potentially developing self-assembly and directed-assembly based materials \cite{Han2017December,*Cheung2017December}. Many techniques that rely on external fields such as optical tweezers \cite{Juan2011,*Marago2013nov}, magnetic tweezers \cite{Wu2017}, thermal ratchets \cite{Magnasco1993September,*Wu2016}, etc. have been developed. However passive systems that require no external input are much more efficient and fundamentally interesting for the development of complex lab-on-a-chip systems. Casimir forces arising from \replaced{quantum vacuum}{charge} fluctuations \cite{Dalvit2011,*Rodriguez2011a} are entirely internal to the system of interest and are thus an attractive \replaced{candidate}{consideration} for developing passive ``nanoparticle ramps''.

The Casimir force has been experimentally measured to be consistent with theoretical predictions \cite{Mohideen1998}, and demonstrated in quantum actuation \cite{Chan2001a} to \replaced{drive}{pump} contactless nano-devices. Lateral Casimir forces that can affect fly-by nanoparticles have also been proposed recently, for a spinning particle near a plate (rotation-induced mirror-symmetry breaking) \cite{Manjavacas2017Mar} and for an anisotropic particle near a plate in thermal nonequilibrium (anisotropy induced) \cite{Mueller2016Mar}. To transport nanoparticles, contact friction from the substrate must be avoided through quantum levitation which, according to the Dzyaloshinski\v{\i}-Lifshitz-Pitaevski\v{\i} (DLP) constraint on the permittivities of the components \cite{Munday2009,Rodriguez2013}, usually needs fluidic environments \replaced{where}{that are incompatible with} \replaced{previously reported mechanisms to generate}{approaches to} lateral Casimir forces \added{do not work} \cite{[{Lateral Casimir forces arising from displacement-induced mirror-symmetry breaking cannot lead to net transport of nanoparticles, but can serve in a time-correlated ratchet, see,~}]Emig2007a}.

\begin{figure}[tb]
\scalebox{0.68}{\includegraphics{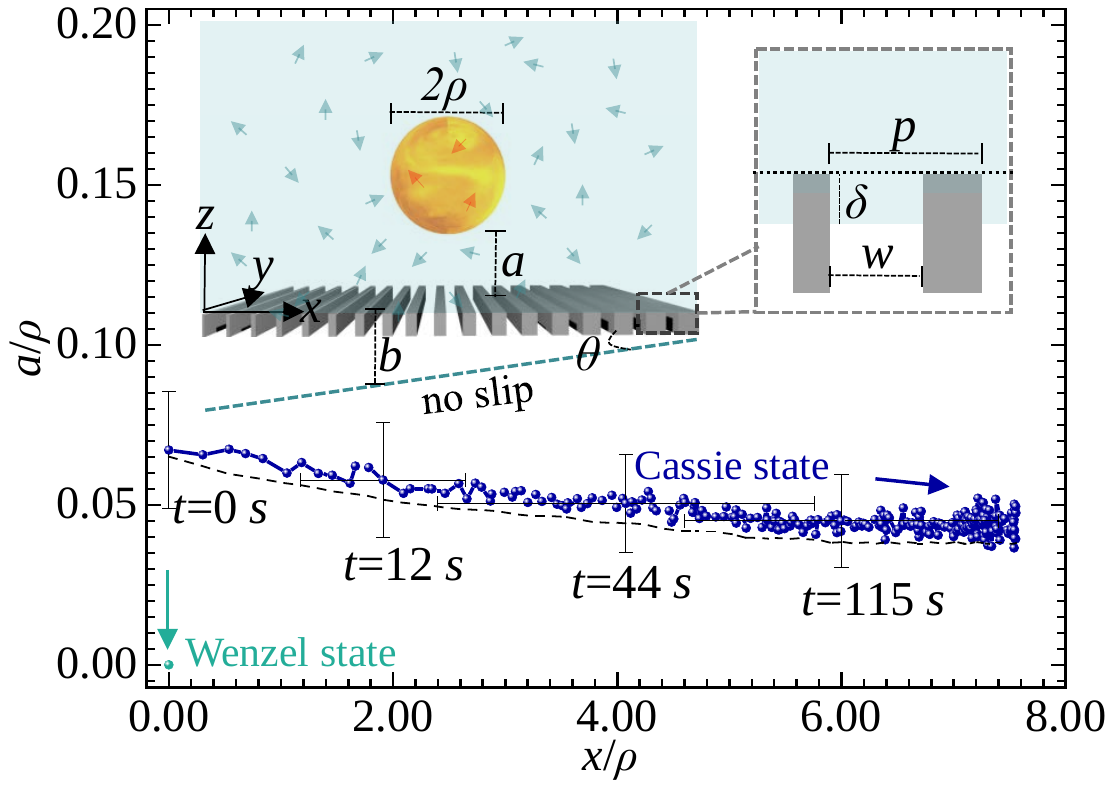}}
\caption{\label{fig:setup} Dynamics of a golden sphere ($ \rho = 1\,\mathrm{\mu m} $) immersed in water above a one-dimensional silica grating ($ f_\mathrm{min}=0.001 $, $ f_\mathrm{max}=0.5 $, $ L/\rho=7.6 $, $ w/\rho=0.1 $) for Cassie ($ \delta/w=0 $) and Wenzel states ($ \delta/w=\infty $) at room temperature ($ T=300\,\mathrm{K} $). The dashed cyan curve below gratings and the dashed black curve below the trajectory represent the effective no-slip boundary and vertical equilibrium heights, respectively, for the Cassie state.}
\end{figure}

This letter proposes inhomogeneity-induced lateral Casimir forces, based on superhydrophobic gradient metasurfaces \cite{[{See, \eg,~}][{ for super-hydrophobicity; }]Liu2014,*[{see, \eg,~}][{ for a review of gradient metasurface.}]Ding2017} as schematically shown in \fref{fig:setup},
and predicts the transport of an immersed nanoparticle driven \deleted{solely }by \replaced{Casimir and/or Langevin stochastic forces}{charge and molecular fluctuations}. This transport process would generally be interrupted if either type of fluctuations is turned off in the Langevin equation, where \replaced{quantum vacuum}{charge} fluctuations generate a washboard-type Casimir energy ramp, while \replaced{stochastic forces}{molecular fluctuations} assist transitions of the nanoparticle to lower-energy positions across energy barriers. This behavior resembles a Brownian motor \cite{Astumian2002,*Haenggi2009Mar}, and no external field is needed.\added{ For generic nanoparticles and gradient metasurfaces, Casimir energy barriers and directional lateral Casimir forces compete, yielding nontrivial transport velocity dependence on various parameters.}\deleted{Superhydrophobicity stemming from nanostructures on metasurfaces leads to mixing liquid-air and liquid-solid interfaces, renders the permittivity of the metasurface effectively smaller, and relaxes the conventional DLP constraint. The proposed system achieves Casimir transport, due to their special responses to both types of fluctuations. Charge fluctuations generate fluctuating electromagnetic (EM) fields with zero mean $ \expv{E_i(\mathbf{r})}=0 $ but nontrivial spatial correlations $ \expv{E_i(\mathbf{r})E_j(\mathbf{r'})}\neq 0 $. Superhydrophobicity stemming from nanostructures on metasurfaces leads to mixing liquid-air and liquid-solid interfacial boundary conditions to both fluidic flows and fluctuating EM fields. Gradient metasurfaces break the mirror symmetry of the system in two length scales, a unit cell of dimension $ p $ and a super cell (featuring gradients) of dimension. The latter constitutes a lateral inhomogeneity and accounts for the proposed lateral Casimir force. Mixing boundary conditions relax the DLP constraint and allow levitation of a nanoparticle for otherwise impossible material configurations. However an object suspended in close proximity to a plate, albeit contactless, is hardly movable either, due to the wall-induced suppression of the hydrodynamic mobility $ \mu $. Mixing boundary conditions further greatly relieve the suppression through a nonzero effective slip length $ b $. Liquid-air menisci, modelled as the sinking depth $ \delta $ and linked to the pressure difference $ \Delta $ of the liquid and air through the Young-Laplace equation $ \delta\sim 2\Delta\cdot w^2/\sigma $ ($ \sigma $ the liquid-air surface tension), significantly impact the Casimir force (both sign and strength) and provide possibilities to real-time modulation.}

Considering a sphere of radius $ \rho $ above a one-dimensional grating of filling factor
\begin{equation}
f(x) = \left\lbrace
\begin{array}{ll}
f_\mathrm{min}, & \mbox{for } x \le 0 \\
f_\mathrm{min} + \frac{x(f_\mathrm{max}-f_\mathrm{min})}{L}, & \mbox{for } 0<x<L \\
f_\mathrm{max}, & \mbox{for } x \ge L
\end{array}
\right.
\end{equation}
where $ L $ is the length of a typical nanoparticle channel. The width of grooves of the grating $ w $ is kept constant, so that the sinking depth of the liquid-air menisci $ \delta $ can be treated identically (period of unit cells is $ p(x)=w/(1-f(x)) $). The Casimir energy of this system at thermal equilibrium, \added{in contrast to the well-known trace-log formula, evaluates the log operation and exactly reads} \footnote{For the derivation, see supplementary materials.}\deleted{
given by the $ \calT $-operator,}
\deleted{where $ \mathcal{I} $ is the unit operator and $ \calG $ is the free photon propagator. For a golden sphere of radius $ \rho $ in water, which scatters plane EM waves into continuous spectra in k-space, the energy can be rewritten as}
\deleted{bearing in mind that the reflection operator of the plate (object), $ \calR_{\mathrm{p(o)}} = -\calG\calT_{\mathrm{p(o)}} $, could be extracted from the Lippmann-Schwinger equation with appropriate translational transformation $ \calX=\exp\{-\sqrt{|\hat{\mathbf{k}}|^2-\e_\mathrm{f}(\xi_n/c)^2}a\} $, and that $ \calT \equiv [\mathcal{I}+\calG\calV]^{-1}\calV $ and $ \calV\equiv -(\tilde{\e}-\tilde{\e}_\mathrm{f})(\xi_n/c)^2 $.}
\begin{equation}\label{eq:energy}
\calE=\frac{-1}{\beta}\sideset{}{'}\sum\limits_{n=0}^\infty\sum\limits_{\gamma}\int\limits_{-\infty}^\infty \langle \mathbf{k}\gamma,\mathrm{in} | \bR_\mathrm{p}\bR_\mathrm{m} | \mathbf{k}\gamma,\mathrm{in} \rangle_n \,\df^2 \mathbf{k},
\end{equation}
where \added{$ \beta=1/k_BT $ ($ k_B $ the Boltzmann constant), $ \ket{\bk\gamma,\mathrm{in}}_n $ is the plane-wave state at a given Matsubara frequency $ \im\xi_n\equiv2\pi n\im/\hbar\beta $ ($ \hbar $ the reduced Planck constant)}, $ \mathbf{k} $ is the lateral wave vector in the $ x $-$ y $ plane\deleted{($ \hat{\mathbf{k}} $ the promoted operator), $ c $ is the speed of light in vacuum, $ \e $ is the permittivity ($ \e_\mathrm{f} $ for the fluid, and tilde means its Fourier image)}, $ \gamma $ = TE or TM represents polarization, and in(out) means the negative(positive)-$ z $ propagation direction. The prime on the summation over Matsubara frequencies indicates that the $ n=0 $ term is weighted by $ 1/2 $. And $ \bR_\mathrm{p(m)} $ is the reflection operator of the nanoparticle(metasurface) to be evaluated at the $ z=0 $ plane. Casimir forces $ \calF_j=-\p_j\calE $ have similar forms as \eref{eq:energy}, but with $ \bR_{\mathrm{m}} $ replaced by
\begin{equation}
-\p_j\bR_{\mathrm{m}} = \left\lbrace
\begin{array}{ll}
-\im[\hat{\mathbf{k}},\bR_{\mathrm{m}}]_-, & \mbox{for }j=x,y\\
\!\,[ \sqrt{\hat{\mathbf{k}}^2+\e_\mathrm{f}\cdot(\frac{\xi_n}{c})^2},\bR_{\mathrm{m}} ]_+, & \mbox{for }j=z
\end{array}
\right.
\end{equation}
where $ [\cdot]_{-(+)} $ is the anti-commutative (commutative) operator and $ \e(\br,\im\xi_n) $ is the permittivity at the corresponding Matsubara frequency ($ \e_\mathrm{f} $ for the uniform fluid; $ c $ the speed of light in vacuum). \added{For gradient metasurfaces, exact periodicity is lost in the unit-cell lengthscale but remains in the super-cell lengthscale \cite{Ding2017}.} $ \bR_{\mathrm{m}} $ can be evaluated by rigorous coupled-wave analysis \added{\cite{[{}][{. For details about how RCWA is applied to gradient metasurfaces, see supplementary materials.}]Davids2010}}, while $ \bR_{\mathrm{p}} $ can be evaluated by \added{partial wave analysis} \deleted{converting plane waves $ | \mathbf{k}\gamma,+ \rangle_{\xi_n} $ to multipole waves $ | lmQ \rangle_{\xi_n} $ ($ l,m $ are quantum numbers for angular momenta; $ Q $ denotes electric or magnetic multipoles), calculating reflected multipole waves using Mie coefficients $ \mathcal{M} $, and then converting them back to plane waves, $ \langle \mathbf{k}\gamma,- | \calR_\mathrm{o} | \mathbf{k}'\gamma',+ \rangle_{\xi_n} = \sum_{lmQ}\langle \mathbf{k}\gamma,- | lmQ \rangle_{\xi_n}\cdot \mathcal{M}_{lmQ}(\xi_n) \cdot \langle lmQ | \mathbf{k}'\gamma',+ \rangle_{\xi_n} $, where the wave transformations }\cite{Neto2008,*Sah2009,*Durand2009,Messina2015Dec}.

\begin{figure}[tb]
\scalebox{0.68}{\includegraphics{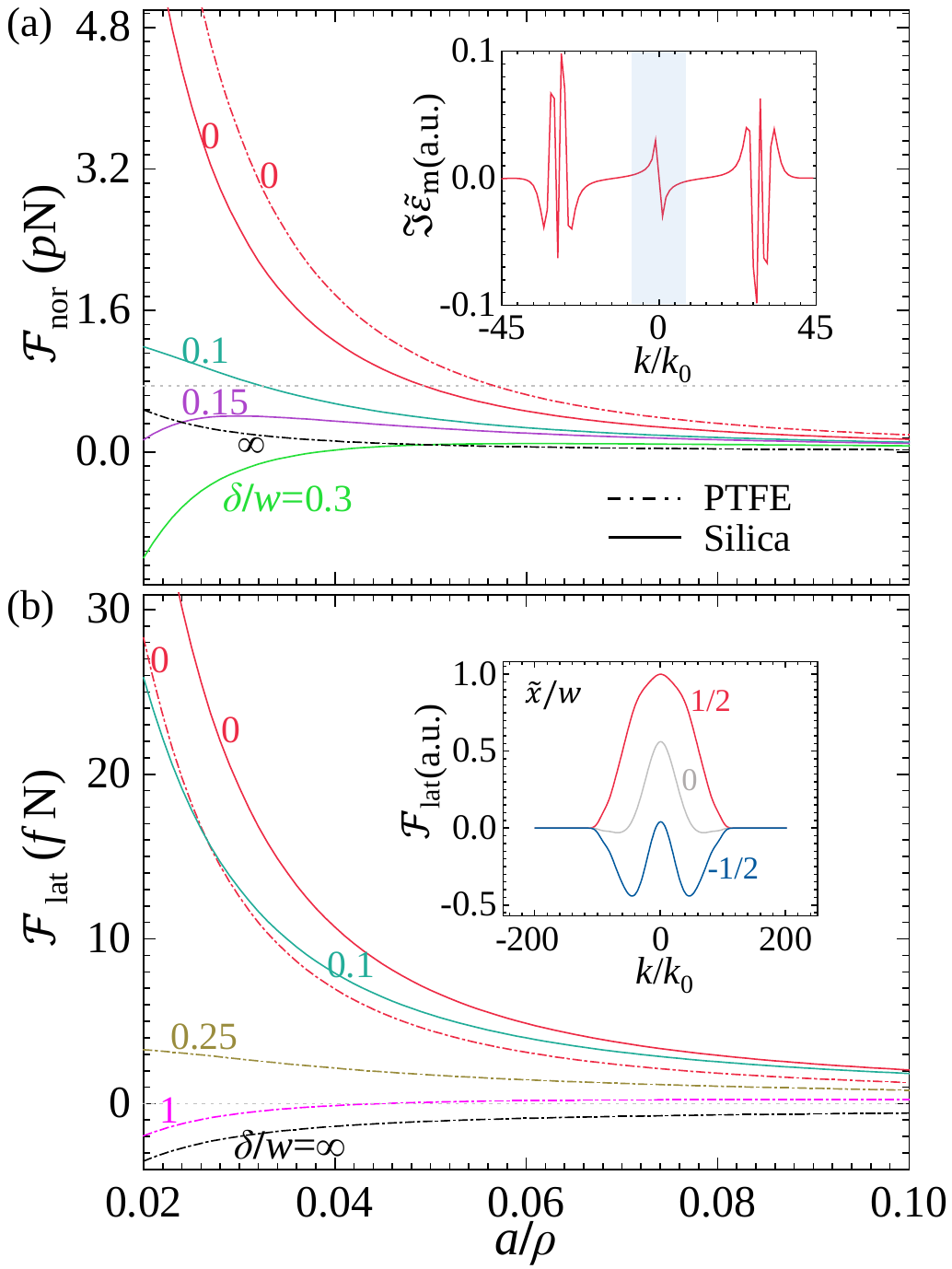}}
\caption{\label{fig:force}(a) Normal and (b) lateral Casimir forces, $ \calF_\mathrm{nor} $ and $ \calF_\mathrm{lat} $, on the sphere described in \fref{fig:setup} (at $ x/\rho=2.405 $, centre of a groove), as functions of particle-metasurface separation $ a/\rho $, for varying sinking depths $ \delta/w $ and substrate materials. $ -G_z=0.75\,p\mathrm{N} $ is also shown in (a) (dotted grey line). \replaced{Top inset: profile of the imaginary part of the Fourier-transformed permittivity of a typical gradient metasurface at Matsubara frequencies. Bottom inset}{Top inset: multipole contributions (of order $ l $) to $ \calF_\mathrm{lat} $. Left bottom inset}: contributions of lateral wave number to $ \calF_\mathrm{lat} $ at different positions ($ \tilde{x}=x-x_0 $, $ x_0/\rho=5.045 $ the centre of a groove, $ a/\rho=0.02 $). $ k_0=2\pi/4.01\,\mathrm{\mu m} $.\deleted{ Right bottom inset: dependence of $ \calF_\mathrm{lat} $ (averaged over local periods) on the filling factor $ f $ at different separations $ a/\rho $.} }
\end{figure}

\fref{fig:force}
shows $ \calF_j $ in the proposed system specified in \fref{fig:setup}, as functions of $ a/\rho $, for varying $ \delta/w $ and substrate materials. \added{In the considered superhydrophobic case, water-silica and water-air interfaces both influence the sphere at a distance of $ a $ and $ a+\delta $, respectively.} For $ \delta/w<0.1 $, the Au-water-silica (silica metasurface) configuration yields repulsive $ \calF_\mathrm{nor} $ which intersects with the $ -G_z $ line ($ G $ represents classical forces including gravity and buoyancy), allowing levitation of the sphere. This also holds for Au-water-silicon and Au-water-Au configurations that have large Hamaker constants (not shown), as long as $ f $ is small enough. For $ \delta/w>0.15 $, $ \calF_\mathrm{nor} $ exhibits non-monotonicity \added{\cite{[{Non-monotonic Casimir forces have been experimentally observed between intersecting nanostructures, see,~}]Tang2017feb}}, with a negative trend in the small-separation limit $ a/\rho\to0 $ where $ (a+\delta)/a $ becomes significant (influence of water-air interfaces significantly weakened as compared with that of water-silica interfaces). Water-PTFE interfaces repel the golden sphere as well as water-air interfaces, so no non-monotonicity shows up when $ \delta/w\to\infty $. \added{From a macroscopic point of view, the above results mean that the total behavior of trapped air and substrate of the metasurface ($ \e_\mathrm{m} $) amounts to that of an effective planar medium ($ 1<\e_\mathrm{eff}<\e_\mathrm{m} $) \cite{[{The valid expression of $ \e_\mathrm{eff} $ in Casimir problems is beyond the scope of this letter. For various effective medium approximations, see, \eg,~}][{}]Azari2010Sep, *Esquivel-Sirvent2011, *Song2017Feb}, and thus the conventional DLP repulsion constraint $ \e_\mathrm{m}<\e_\mathrm{f}<\e_\mathrm{p} $ is significantly relaxed to $ \e_\mathrm{eff}<\e_\mathrm{f}<\e_\mathrm{p} $. For gradient metasurfaces, mirror-symmetry breaking in the unit-cell lengthscale (captured by side peaks in the top inset) is common in previously studied periodic gratings, while symmetry breaking in the super-cell lengthscale (shadowed peaks around $ \pm k_0 $ \footnote{$ k_0 $ comes from the re-construction technique. See supplementary materials Sec. SII.}) generates lateral inhomogeneity \cite{[{The Casimir effect within an inhomogeneous system has been investigated previously, see, \eg,~}]Bao2015,*Bao2016} that accounts for the inhomogeneity-induced lateral Casimir force. $ \calF_{\mathrm{lat}} $ of magnitudes comparable with the particle's weight can also be analyzed via competing contributions from different interfaces. Notably, at a fixed particle position, $ \calF_{\mathrm{lat}} $ from the PTFE metasurface flips sign for increasing $ \delta $, while $ \calF_{\mathrm{lat}} $ from the silica metasurface remains directional.} The Au permittivity in computations is obtained from a Drude model, $ \e_\mathrm{Au}=1\replaced{+}{-} \Omega^2/\xi_n(\xi_n+\deleted{\im}\Gamma) $, with plasma frequency $ \Omega=1.28\times10^{16}\,\mathrm{rad/s} $ and damping constant $ \Gamma=6.60\times10^{13}\,\mathrm{rad/s} $. The silica permittivity is fitted by Lorentz terms from tabular data \cite{Palik1985}. Permittivities of water and polytetraﬂuoroethylene (PTFE) are obtained from Lorentz models with parameters given in Ref.~\cite{Zwol2010}. All magnetic responses are ignored.\deleted{ 401 orders of spatial harmonics are used in RCWA to decompose the EM fields and permittivities. $ 100 $ orders of spherical harmonics and $ n\le56 $ prove sufficient to yield convergent results.}

These new features of Casimir forces lead to the ensemble-averaged trajectory ($ \expv{a(t)}/\rho $ versus $ \expv{x(t)}/\rho $) of the sphere in \fref{fig:setup}, according to the Langevin equation
\begin{equation}\label{eq:langevinEq}
\dot{r}_i=\mu_{ij}\left[ \calF_j + G_j - m\ddot{r}_j \right] + \nu_{ij}\zeta_j(t),
\end{equation}
where $ \nu_{ij}\nu_{jk}= 2\mu_{ik}/\beta $ is implied by the fluctuation-dissipation theorem ($ \mu_{ij} $ the position-dependent mobility tensor), $ -m\ddot{r} $ is the inertial force (dot means derivative with respect to time), and $ \zeta $ with zero mean $ \expv{\zeta_i(t)}=0 $ and temporal correlation $ \langle\zeta_i(t)\zeta_j(t')\rangle=\delta_{ij}\delta(t-t') $ represents Gaussian white noise. The effective slip length of this system for the Cassie state ($ \delta/w=0 $)
\begin{equation}
b(x) = \frac{p(x)}{2\pi}\ln\sec[\frac{\pi}{2}(1-f(x))],
\end{equation}
in comparison to uniform gratings \cite{BELYAEV2010}, implies that the effective no-slip boundary is no longer parallel to the $ z=0 $ plane due to gradients, but with an angle $ \theta = \arctan\p_x b $, and thus the mobility tensor generally features off-diagonal terms $ \mu_{xz} $ and $ \mu_{zx} $. \deleted{For lateral transports where $ \expv{\dot{a}}\ll\expv{\dot{x}} $, the effective lateral mobility $ \mu_\mathrm{eff}=\mu_{xx}(1-\mu_{xz}\mu_{zx}/\mu_{xx}\mu_{zz}) $, derived by setting $ \dot{a}=0 $, is thus smaller than the apparent lateral mobility $ \mu_{xx} $, due to the coupling of lateral and normal motions. }In the present case, $ \theta $ is found to be within 4 degrees for which \replaced{off-digonal terms $ \sim\sin\theta $}{the factor $ \mu_{xz}\mu_{zx}/\mu_{xx}\mu_{zz} \sim \sin^2\theta $} are vanishingly small, and thus lateral and normal motions of the sphere decouple. \added{It turns out that different mobility profiles of $ \mu_\parallel $ and $ \mu_\perp $ do not alter our conclusions} \footnote{See supplementary materials Sec. SIII.}, and it is sufficient to model them as
\begin{multline}\label{eq:lateralMu}
\mu_\parallel = \mu_0 (1-\frac{9}{16}\tau_\parallel+\frac{1}{8}\tau_\parallel^3-\frac{45}{256}\tau_\parallel^4-\frac{1}{16}\tau_\parallel^5-\frac{83}{256}\tau_\parallel^6),\\
\mu_\perp = \mu_0 (1-\frac{9}{8}\tau_\perp+\frac{1}{2}\tau_\perp^3-0.535\tau_\perp^4+0.160\tau_\perp^5),
\end{multline}
where $ \tau_\perp = \rho/(\rho+a) $, $ \tau_\parallel = \rho/(\rho+a+b) $, and $ \mu_0 $ is the bulk mobility. Here it is further required that $ \mu\to 0 $ when $ \tau \to 1 $.
The Cassie-state trajectory, obtained for 100 repeated finite-difference simulations for a total time of 270 s at a time step of $ \delta t=0.01\,\mathrm{s} $ \footnote{The trajectory is shown with $ \delta t=1\,\mathrm{s} $ rather than $ 0.01\,\mathrm{s} $, where the transport velocity can be revealed from the data density.}, illustrates the transport behavior with an average speed of about $ 2 \,\mathrm{\mu m/s} $.\deleted{, and shown with $ \delta t=1\,\mathrm{s} $. Standard deviations $ \Delta x(t_i) $ (only shown for 4 different moments) exhibit non-monotonic dependence with respect to time where $ \Delta x(t=115\,\mathrm{s}) $ gets smaller as compared with that at $ t= 44\,\mathrm{s} $, due to the existence of energy barriers for $ x/\rho>4 $ (not shown). Spatial separations $ |\expv{x(t_i)}-\expv{x(t_j)}| $ of these moments are beyond the deviations, indicating that this transport over about 4 body-size in $ x $-direction is significantly observable. Initial conditions are set as $ x(t=0\,\mathrm{s})/\rho=0 $ and $ a(t=0\,\mathrm{s})/\rho $ a random value between 0.04 and 0.1. Fixed initial height $ a(t=0\,\mathrm{s})/\rho = 0.067 $ does not change other statistical features, implying the considered case is in the overdamped region.} The Wenzel-state trajectory, on the other hand, drops immediately to around $ a/\rho = 0 $ and $ x/\rho = 0 $ from the beginning, as well as the case ignoring $ \calF_j $.\deleted{ Once charge fluctuations are turned off, \ie, ignoring the term $ -\p_j\calE $, the sphere also drops to the origin, as expected.} If \replaced{stochastic forces}{molecular fluctuations} are turned off, the sphere (Cassie state) is found to travel along the equilibrium-height (dashed black) curve but get stopped halfway by energy barriers (not shown). Diffusion makes $ \expv{a}\equiv\int a\exp\{-\beta\calE_\mathrm{tot}\}\,\df a/\int \exp\{-\beta\calE_\mathrm{tot}\}\,\df a $ larger than the equilibrium height where $ \calF_\mathrm{nor}+G_z=0 $, due to the asymmetry of $ \calE_\mathrm{tot}(a)=\calE(a)-G_z\cdot a $ along $ z $-direction. This quantitatively explains the fact that the Cassie-state trajectory in \fref{fig:setup} is $ 5.8\,\mathrm{nm} $ above the equilibrium height. \added{Tuning the sinking depth $ \delta $ via pressure difference $ \Delta $ of the liquid and air through the Young-Laplace equation $ \delta\sim 2\Delta\cdot w^2/\sigma $ ($ \sigma $ the liquid-air surface tension) significantly affects the Casimir force (both sign and strength, see \fref{fig:force}), and thus enables on-off or directional switching of the transport process.}

More generally, mirror-symmetry breaking of the one-dimensional-gradient (along the $ x $-axis) metasurface could be simplified as (top inset of \fref{fig:force})
\begin{equation}\label{eq:msb}
\begin{split}
\Im\tilde\e_\mathrm{m}(k_x,\im\xi_n) \approx {}
& -\added{{q}_1} \mathrm{sgn}(k_x)\delta(|k_x|-\frac{2\pi}{p}) \\
& - \added{{q}_2} \mathrm{sgn}(k_x)\delta(|k_x|-\frac{2\pi}{P}) ,
\end{split}
\end{equation}
where $ P $ is the super-cell dimension and $ \Im\tilde{\e} $ is the imaginary part of the Fourier transformed (in $ x $-axis; $ y $ and $ z $ dependences suppressed) permittivity. According to the Kramers-Kronig relation $ \e_\mathrm{m}(\mathbf{r},\im\xi_n)=1+\frac{2}{\pi}\int_{0}^{\infty}\omega\Im\e_\mathrm{m}(\mathbf{r},\omega)/(\omega^2+\xi_n^2)\,\df\omega $, $ {q}_1 $ and $ {q}_2 $ would always be zero if the mirror symmetry holds. \eref{eq:msb} corresponds to sinusoidally modulated permittivities, and under the Born approximation entails \cite{[{Derivations are quite similar to the case of sinusoidally corrugated surfaces, see, \eg,~}][]Dalvit2008}
\begin{equation}\label{eq:bornApprxm}
\calF_{\mathrm{lat}} = \added{q_1}A\cos(\frac{2\pi}{p}x) + \added{q_2}B
\end{equation}
in the $ P\gg x $ limit, which implies that the Casimir energy
\begin{equation}\label{eq:bornEnergy}
\calE(x) = -\added{q_1}\frac{Ap}{2\pi}\sin(\frac{2\pi}{p}x) - \added{q_2}Bx - \added{C}
\end{equation}
resembles a washboard-type potential ramp for the nanoparticle. Here, properties of the particle play their roles through $ A $ and $ B $. The second term of \eref{eq:bornApprxm} is the inhomogeneity-induced directional force, while the first term oscillates with $ x $, qualitatively consistent with the bottom inset in \fref{fig:force} ($ P=\tilde{P}=4.01\,\mathrm{\mu m} $).

\begin{figure}[tb]
\scalebox{0.6}{\includegraphics{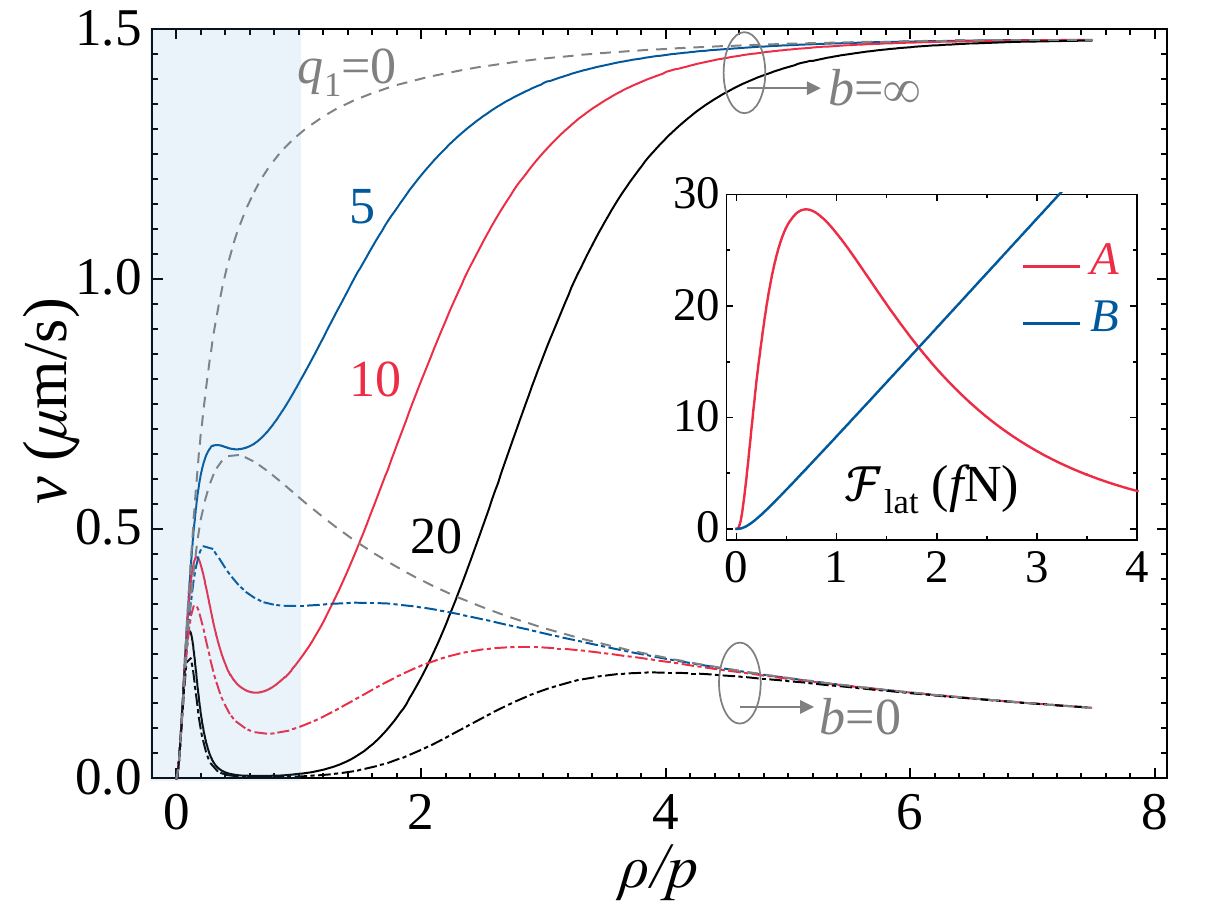}}
\caption{\label{fig:velocityR}\added{Room-temperature Casimir transport velocity $ v $ as a function of radius $ \rho $ (normalized by $ p=200\,\mathrm{nm} $), for $ a=50\,\mathrm{nm} $, $ P=16.08\,\mathrm{\mu m} $ and $ q_2=1 $. Inset: $ A $ and $ B $ as functions of $ \rho/p $.} }
\end{figure}

\added{\fref{fig:velocityR}}
\added{shows $ v\equiv\dot{x} $ as a function of $ \rho/p $, according to \eref{eq:bornEnergy} and the overdamped Stratonovich formula \cite{Lindner2001}
\begin{equation}\label{eq:stratonovich}
v = \frac{p\mu_\parallel(1-e^{-\beta q_2Bp})/\beta}{\int_{0}^{p}\df x e^{-\beta\calE(x)}\int_{x}^{x+p}\df y e^{\beta\calE(y)} }.
\end{equation}
When $ q_2=0 $, \eref{eq:stratonovich} yields $ v=0 $ and the proposed system recovers the interaction between a sphere and a periodic grating where Casimir transport does not occur. The asymptotic ($ q_1=0 $) velocity (dashed grey) in the $ b=\infty $ limit, decreases rapidly when $ \rho/p<1 $ and reaches a plateau when $ \rho/p>1 $, due to the behavior of $ B/\rho $. $ B(\rho) $ (inset) is almost linear for large particle size, while $ B=\mathcal{O}(\rho) $ when $ \rho\to0 $. For nontrivial energy barriers (large $ q_1 $), the transport velocity shows a sharp peak in the $ \rho/p<1 $ region (shadowed), increases rapidly in the $ 1<\rho/p<5 $ range, and approaches the asymptotic plateau when $ \rho/p>5 $, due to the behavior of $ A $. For $ \rho/p<0.5 $, both $ A $ and $ B $ increase with $ \rho $, but $ A $ is much faster. For $ \rho/p>1 $, force contributions from neighboring unit cells compete with each other and thus $ A $ decreases, leaving a peak around $ 0.5<\rho/p<1 $, while $ B $ still increases. The nanoparticle is in the running state most of the time when $ q_1A $ is small, while it gets locked by energy barriers most of the time when $ q_1A $ becomes large. Therefore, the effective transport velocity is low in the region where $ A $ is large, and significantly deviates from the asymptotic curve. With increasing $ \rho/p $ the asymptotic velocity in the $ b=0 $ limit increases rapidly at first but slowly decays later, due to the suppression of $ \mu_{\parallel} $ by wall-induced hydrodynamic interactions (\eref{eq:lateralMu}). Nontrivial energy barriers again result in sharp peaks of the velocity in the $ \rho/p<1 $ region, indicating that the size-selective transport behavior is robust against the slippage of the metasurface. In all computations, $ A $ and $ B $ are generated assuming perfect conducting boundaries on the particle, for simplicity.}

\begin{figure}[tb]
\scalebox{0.6}{\includegraphics{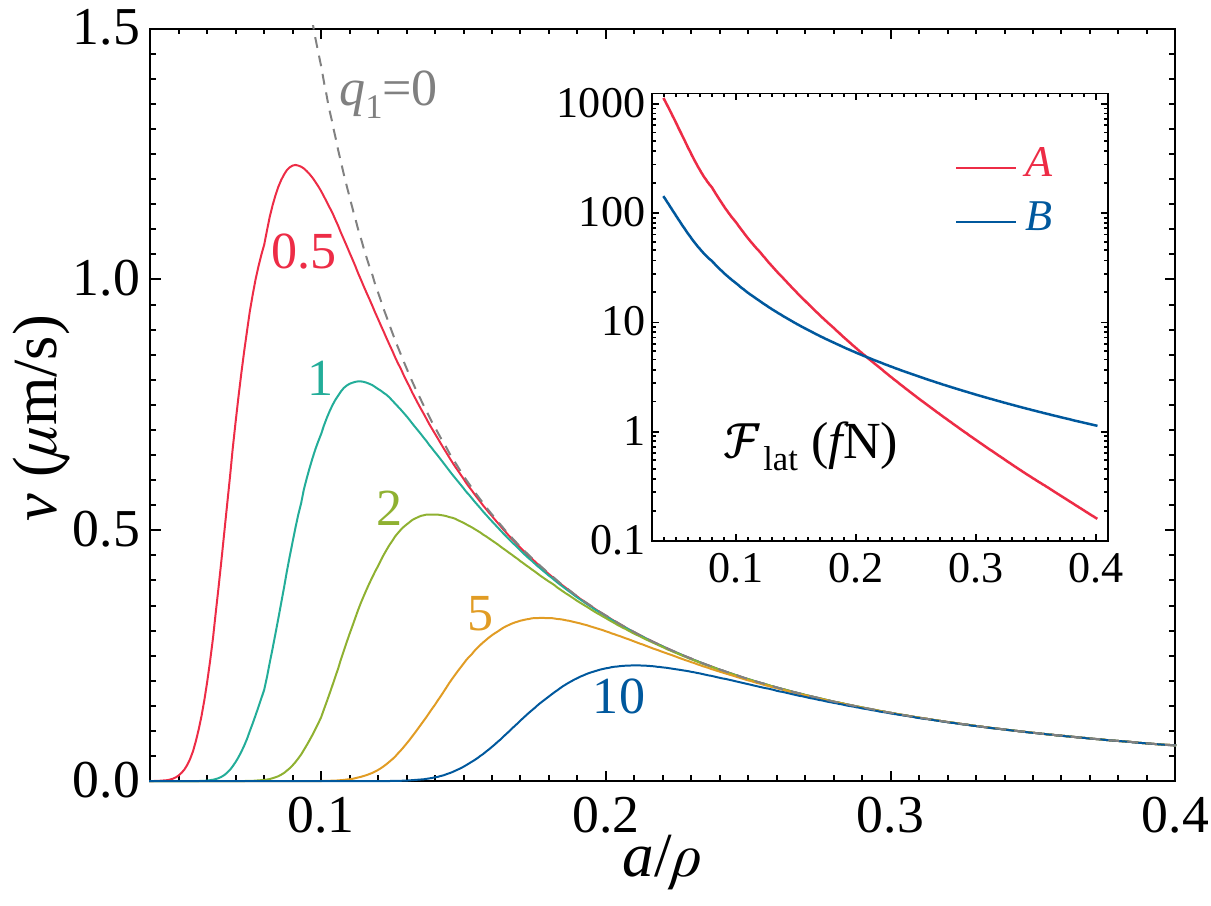}}
\caption{\label{fig:velocityA}\added{Room-temperature Casimir transport velocity $ v $ as a function of separation $ a $ (normalized by $ \rho=500\,\mathrm{nm} $), for $ b=\infty $, $ q_2=1 $, $ p=400\,\mathrm{nm} $, and $ P=16.08\,\mathrm{\mu m} $. Inset: $ A $ and $ B $ as functions of $ a/\rho $.} }
\end{figure}

\added{Competition between energy barriers (term $ A $) and directional forces ($ B $) results in an optimal value of $ a/\rho $ (\fref{fig:velocityA}).}
\added{The asymptotic velocity (dashed grey) in the $ b=\infty $ limit diverges when $ a/\rho\to0 $ and decays when $ a/\rho\to\infty $, similar with the behavior of the Casimir force $ B(a) $. With any finite energy barrier, $ v\to0 $ when $ a/\rho\to0 $. As is shown in the inset, $ A $ diverges faster than $ B $ in the $ a/\rho\to0 $ limit, and decays faster in the opposite limit. This means that, for any nonzero $ q_1 $, a finite $ a/\rho $ must exist below which energy barriers begin to dominate. The separation range around the velocity peak defines a priority passage for the nanoparticle, outside which nanoparticles are transported much slower, where smaller energy barriers result in narrower passages. The peaks are not around where $ A\le B $, so energy barriers are still important when $ v $ is optimized. The $ b\to0 $ limit yields similar curves (not shown), indicating that those features are also robust against the slippage of the metasurface.}

\added{When the temperature of the system varies, Matsubara frequencies change and Casimir forces (both $ A $ and $ B $) modestly increase with $ T $ (not shown). The total influence, according to \eref{eq:stratonovich}, is that $ v\propto 1/\beta\propto T $, which might be another way to control the transport.}

In the proposed systems, Lewis acid-base interactions and electrostatic double-layer interactions are also present \cite{Oss2003}. The former is usually within a range of $ 3\,\mathrm{nm} $ away from the plate, and the latter can be suppressed by using uncharged surfaces or tuning the Debye length to a similar range. Experiments have demonstrated pure Casimir effects without influence from those two interactions when the particle-metasurface separation is beyond $ 10\,\mathrm{nm} $ \cite{Munday2009,Tabor2011}, and thus allow verification of the above results. \added{The inhomogeneity-induced lateral Casimir force can also affect fly-by nanoparticles without fluidic environments. The Casimir transport and its velocity's dependence open new opportunities for developing technologies and explaining fundamental physical, biological or chemical processes at the nanoscale.}
\begin{acknowledgments}
This work was partially supported by China Postdoctoral Science Foundation (Grant No. 2017M622722), the National Key Research and Development Program of China (No. 2017YFA0205700) and the National Natural Science Foundation of China (No. 11621101). The authors thank the anonymous referees for helpful comments.
\end{acknowledgments}

\bibliography{Casimir,NanoDynamics,meta,TO}
\end{document}


\title{Supplementary Materials: Inhomogeneity-Induced Casimir Transport of Nanoparticles}
\author{Fanglin Bao}
\affiliation{Centre for Optical and Electromagnetic Research, Guangdong Provincial Key Laboratory of Optical Information Materials and Technology, South China Academy of Advanced Optoelectronics, South China Normal University, Guangzhou 510006, China}
\author{Kezhang Shi}
\affiliation{Centre for Optical and Electromagnetic Research, JORCEP, Zhejiang University, Hangzhou 310058, China}
\author{Guanjun Cao}
\affiliation{Centre for Optical and Electromagnetic Research, Guangdong Provincial Key Laboratory of Optical Information Materials and Technology, South China Academy of Advanced Optoelectronics, South China Normal University, Guangzhou 510006, China}
\author{Julian S. Evans}
\affiliation{Centre for Optical and Electromagnetic Research, JORCEP, Zhejiang University, Hangzhou 310058, China}
\author{Sailing He}
\affiliation{Centre for Optical and Electromagnetic Research, Guangdong Provincial Key Laboratory of Optical Information Materials and Technology, South China Academy of Advanced Optoelectronics, South China Normal University, Guangzhou 510006, China}
\affiliation{Centre for Optical and Electromagnetic Research, JORCEP, Zhejiang University, Hangzhou 310058, China}
\affiliation{Department of Electromagnetic Engineering, Royal Institute of Technology, 10044 Stockholm, Sweden}

\maketitle

\section{SI. Casimir energy}
Here we derive the Casimir energy of Eq.~(2) in the main text for a particle-metasurface (p-m) system, by the transition $ \bT $-operator approach. At thermal equilibrium of temperature $ T $, the Casimir energy takes a general form \cite{Kenneth2006,*Kenneth2008},
\begin{equation}\label{eq:TGTG}
\calE=\frac{1}{\beta}\sideset{}{'}\sum\limits_{n=0}^\infty\ln\mathrm{det}[\bI-\bT_\mathrm{p} \bG \bT_\mathrm{m} \bG],
\end{equation}
where $ \beta=1/k_BT $, $ k_B $ is the Boltzmann constant, $ \bI $ is the unit operator, $ \bG $ is the free photon propagator, and the prime on the summation over Matsubara frequencies $ \im\xi_n=\im\cdot 2\pi n/\hbar\beta $ ($ \hbar $ the reduced Planck constant) indicates that the $ n=0 $ term is weighted by $ 1/2 $. \eref{eq:TGTG} is derived and holds on \emph{discrete} state spaces, and the transition $ \bT $-operator is defined by a potential operator $ \bV $ \cite{Sah2009},
\begin{equation}\label{eq:T}
\bT_\mathrm{p(m)} \equiv \bV_\mathrm{p(m)}[\bI+\bG\bV_\mathrm{p(m)}]^{-1}.
\end{equation}
In the presence of the potential, a free/input field $ |\psi\rangle $ yields a total scattered/output field $ |\Psi\rangle $ that, according to the Lippmann-Schwinger equation \cite{Lippmann1950}, obeys $ |\Psi\rangle = [\bI+\bG\bV]^{-1} |\psi\rangle \equiv (\bI+\bR)|\psi\rangle $ which entails the reflection operator $ \bR_\mathrm{p(m)}=-\bG\bT_\mathrm{p(m)} $. Therefore, we arrive at
\begin{equation}\label{eq:RR}
\calE=\frac{1}{\beta}\sideset{}{'}\sum\limits_{n=0}^\infty\tr\ln[\bI-\bR_\mathrm{p} \bR_\mathrm{m}],
\end{equation}
bearing in mind that two reflection operators should be evaluated in the same reference frame.

For a generic p-m system, where the top of the metasurface lies in the $ z=0 $ plane and laterally extends to infinity, with the particle placed above the metasurface, we follow the box-normalization technique (in the $ x $-$ y $ plane) and denote the lateral dimension of the system as $ 2\mathcal{L} $. Under periodic boundary conditions, the plane-wave basis $ | \mathbf{k}\gamma, s \rangle_n $ is discrete and the lateral wave vector is $ \mathbf{k}_{ij}=\hat{\mathbf{x}}\pi i/\mathcal{L}+\hat{\mathbf{y}}\pi j/\mathcal{L} $ (the normal wave number $ k_z=s\sqrt{\bk^2+\e_\mathrm{f}\cdot(\xi_n/c)^2} $, $ c $ is the speed of light in vacuum, $ \e_\mathrm{f}(\br,\im\xi_n) $ is the permittivity for the uniform fluid), where $ \gamma $ represents the polarization, and $ s=\im(-\im) $ is denoted as $ s=\mathrm{out(in)} $. We will take the continuum limit $ \mathcal{L}\to\infty $, $ \delta k=\pi/\mathcal{L}\to 0 $ to recover results of an unbounded p-m system, and we have
\begin{equation}\label{eq:limit}
\begin{split}
\calE_n 
&\equiv \tr\ln\left[\bI-\bR_\mathrm{p}\bR_\mathrm{m}\right]_n \\
&=\lim\limits_{\delta k\to 0}\sum\limits_{\gamma}\sum\limits_{\bk}\ln\left[ 1-\langle \mathbf{k} \gamma, \mathrm{in} | \bR_\mathrm{p}\bR_\mathrm{m} | \mathbf{k} \gamma, \mathrm{in} \rangle_n \right].
\end{split}
\end{equation}
We can further show that in the $ \ln[1-\mathcal{M}] $ type expression in the above equation $ \mathcal{M} $ is arbitrarily small in the continuum limit, so that $ \ln[1-\mathcal{M}]=-\mathcal{M} $ exactly holds. $ \bra{\bk\gamma, s} \bR_\mathrm{p} \ket{\bk'\gamma', s}_n $ is proportional to the on-shell elements of $ \bra{\bk k_z} \bV_{\mathrm{p}} \ket{\bk' k'_z} $ which can be obtained by Fourier transformation of $ \bra{\br}\bV_{\mathrm{p}}\ket{\br}_n=\delta\e(\br,\im\xi_n)\cdot\left( \frac{\xi_n}{c} \right)^2 $, where the dielectric contrast $ \delta\e=\e_\mathrm{p}-\e_\mathrm{f} $ is finite within the particle (of radius $ \rho $) and vanishes elsewhere. Explicitly, we have
\begin{equation}
\begin{split}
\bra{\bk, z} \bV_{\mathrm{p}} \ket{\bk', z}_n = {}
& \frac{1}{4\mathcal{L}^2}\iint\limits_{-\rho}^{\mathcal{\rho}}\delta\e(\br,\xi_n)\cdot\left( \frac{\xi_n}{c} \right)^2\\
& {}\times \proj{\bk}{x,y}\proj{x,y}{\bk'}\,\df x\df y,
\end{split}
\end{equation}
which is proportional to $ (\rho/\mathcal{L})^2 $, and thus
\begin{equation}
\bra{\bk\gamma, \mathrm{in}} \bR_\mathrm{p} \ket{\bk'\gamma', \mathrm{out}}_n  \propto  (\delta k)^2.
\end{equation}
$ \bra{\bk'\gamma', \mathrm{out}} \bR_\mathrm{m} \ket{\bk\gamma, \mathrm{in}}_n $ is bounded in magnitude when $ \delta k\to0 $. Therefore,
\begin{equation}
\begin{split}
\mathcal{M}={}
&\sum\limits_{\bk' \gamma'} \bra{\bk\gamma,\mathrm{in}} \bR_\mathrm{p} \ket{\bk'\gamma',\mathrm{out}}_n \\
&\times \bra{\bk'\gamma', \mathrm{out}} \bR_\mathrm{m} \ket{\bk\gamma, \mathrm{in}}_n=\mathcal{O}[(\delta k)^2].
\end{split}
\end{equation}
And \eref{eq:limit} reduces to
\begin{equation}\label{eq:nolog}
\calE_n = -\lim\limits_{\delta k\to 0}\sum\limits_{\gamma}\sum\limits_{\bk}\mathcal{M} + \mathcal{O}[(\delta k)^4].
\end{equation}
In the continuum limit where $ \bk $ is \emph{continuous}, transition probabilities become probability densities, and the following substitution holds
\begin{equation}
\langle \mathbf{k} \gamma, \mathrm{in} | \bR_\mathrm{p}\bR_\mathrm{m} | \mathbf{k} \gamma, \mathrm{in} \rangle_n \gets \frac{\mathcal{M}}{(\delta k)^2}.
\end{equation}
Finally, we arrive at
\begin{equation}\label{eq:resultE}
\calE=\frac{-1}{\beta}\sideset{}{'}\sum\limits_{n=0}^\infty\sum\limits_{\gamma}\int\limits_{-\infty}^\infty \langle \mathbf{k} \gamma, \mathrm{in} | \bR_\mathrm{p}\bR_\mathrm{m} | \mathbf{k} \gamma, \mathrm{in} \rangle_n \,\df^2\bk.
\end{equation}

The above derivation relies only on the boundedness of $ \bra{\bk\gamma s} \bR \ket{\bk'\gamma' s'} $ in continuous plane-wave bases, for at least one of the two interacting objects. We can interpret that, after one round trip, an input state has been scattered into a continuous spectrum of states and the probability of reflection into the original state is infinitesimal, so that multiple scatterings are negligible. For planes and gratings that are studied previously, their reflection probability densities contain Dirac delta functions and do not meet the boundedness requirement, and thus \eref{eq:resultE} does not apply. Our results shall be valid not only for Casimir interactions that involve particles, but also for that involve diffused reflectors. Based on \eref{eq:resultE}, we successfully reproduced results of \cite{Neto2008} to verify our $ \bT $-operator codes.

\section{SII. Computations}
This section shows how the reflection operator of a metasurface $ \bR_m $ is computed. For gradient metasurfaces, exact periodicity is lost in the unit-cell lengthscale but remains in the super-cell lengthscale. This enables us to adopt the rigorous coupled-wave analysis (RCWA, or the modal approach) \cite{Davids2010}, which utilizes eigenmodes of the electromagnetic field in various stratified regions and solves scattering amplitudes by matching boundary conditions, to compute $ \bR_m $. In comparison with periodic gratings, higher order Brillouin zones are required in computations, for gradient metasurfaces, to reach convergent results. Since in the main text we require the dimension of the super cell $ P $ be much larger than the dimension of a typical nanoparticle channel $ L $, and $ L $ be much larger than the dimension of the unit cell $ p $, a reconstruction technique is used in practical computations to avoid extremely high order Brillouin zones. For a given metasurface ($ P $), we construct a series of virtual metasurfaces ($ \tilde{P}_j $) which approach the original one asymptotically.

\begin{figure*}[htb]
\scalebox{0.45}{\includegraphics{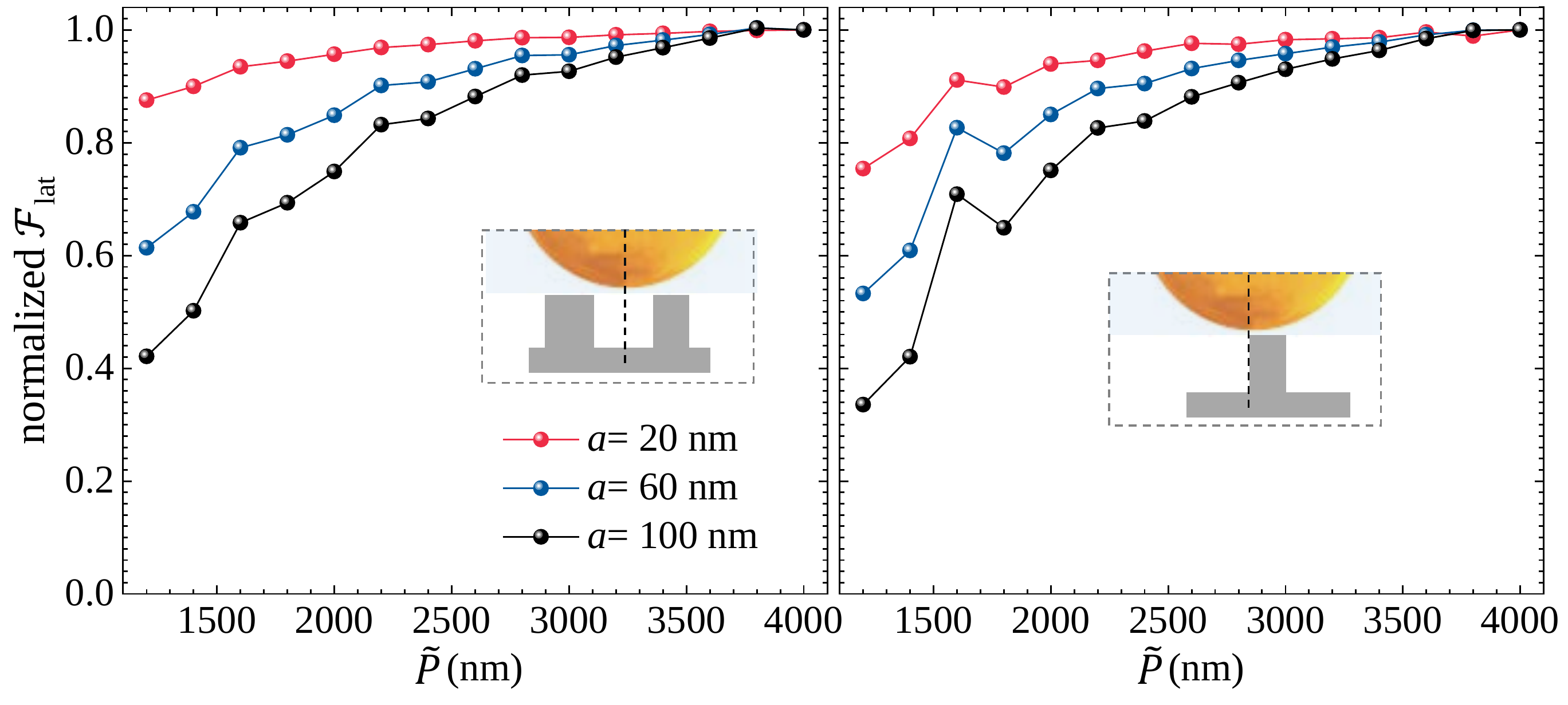}}
\caption{\label{fig:convergence} (color online) Normalized room-temperature lateral Casimir force $ \calF_\mathrm{lat} $ for $ \rho=1\,\mathrm{\mu m} $ and $ f=1/3 $, as a function of the dimension $ \tilde{L} $ of the supercell used in calculations. Left: the center of the sphere aligns with the center of the groove. Right: the center of the sphere aligns with the rising edge. }
\end{figure*}

Explicitly, in a $ j_\mathrm{th} $ ($ j=1,2,3,\cdots $) run of computation, we cut a small patch (of dimension $ \tilde{P}_j $, nearest to the sphere) of the metasurface, and define it as a virtual super cell and periodically duplicate it to construct a virtual gradient metasurface, based on which we can calculate the Casimir force $ \calF_j $ using RCWA. Then in the $ (j+1)_\mathrm{th} $ run, we enlarge $ \tilde{P}_j $ to $ \tilde{P}_{j+1} $ and get $ \calF_{j+1} $, and so on. Eventually we can recover the original metasurface when $ \tilde{P}_J=P $, and obtain the desired force $ \calF $. Since major contributions of the Casimir force exerted on the sphere come from a small area of the metasurface nearest to the sphere, one can expect that $ \calF_j $ converges to $ \calF $ before the $ J_\mathrm{th} $ run. In the case described in the main text, $ L=7.6\,\mathrm{\mu m} $, $ \rho=1\,\mathrm{\mu m} $, and $ P\gg L $. The convergence test in \fref{fig:convergence} shows that $ \tilde{P}=4\,\mathrm{\mu m} $ is sufficient to get the right force (the lateral force converges the slowest; the Casimir energy and the normal force are not shown). $ \tilde{P}=4.01\,\mathrm{\mu m} $ is used to generate force data in the main text ($ k_0\equiv 2\pi/\tilde{P} $), and correspondingly 401 orders of Brillouin zones are sufficient to achieve convergence.

\section{SIII. Hydrodynamic mobility}
\begin{figure}[htb]
\scalebox{1}{\includegraphics{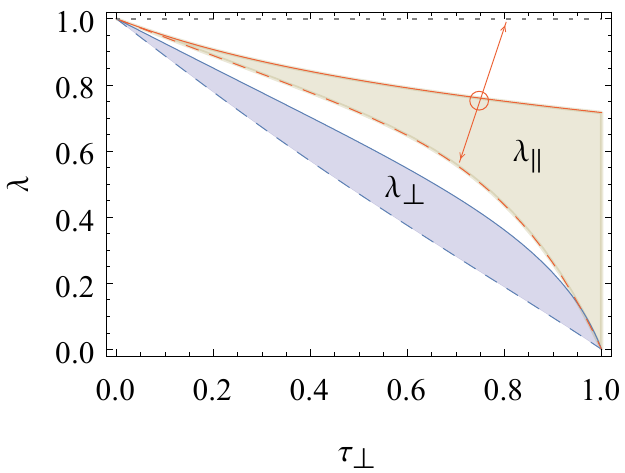}}
\caption{\label{fig:mobilityBound} (color online) The available range of $ \lambda_\perp $ and $ \lambda_\parallel $ for $ b/\rho=1 $, as a function of $ \tau_\perp $. }
\end{figure}

Here we prove that our prediction on the Casimir transport is robust against variations in the hydrodynamic mobility profiles. For laminar flow upon an arbitrary metasurface with a characteristic slip length $ b $, the effective no-slip plane $ z=-b $ (where parallel velocity $ v_\parallel=0 $) and the actual liquid-metasurface interface $ z=0 $ (perpendicular velocity $ v_\perp=0 $) separate. By virtue of Faxen's law, the lateral(normal) mobility $ \mu_\parallel $($ \mu_\perp $) can be constructed with the Green's function of the Stokes equation, under boundary conditions $ v_\perp=0 $ at $ z=0 $ and $ v_\parallel=0 $ at $ z=-b $. The lower(upper) bound of  $ \mu_\perp\equiv\mu_0\lambda_\perp $ is the mobility assuming no(perfect)-slip condition on the interface \cite{Dubov2014}, while the lower(upper) bound of  $ \mu_\parallel\equiv\mu_0\lambda_\parallel $ is the mobility assuming $ v_\parallel(v_\perp)=0 $ on the interface(effective no-slip plane), described in Eqs.~(\ref{eq:inf}-\ref{eq:sup}), where $ \alpha=\mathrm{cosh}^{-1}(1/\tau_\perp) $, $ \tau_\perp=\rho/(\rho+a) $, and $ \tau_\parallel=\rho/(\rho+a+b) $ \cite{Happel1983}. The available range of $ \lambda $ for an arbitrary metasurface is shown in \fref{fig:mobilityBound}. Note that $ \mathrm{Sup}\{ \lambda_\parallel \}\to1 $ when $ b\to\infty $, and it coincides with $ \mathrm{Inf}\{ \lambda_\parallel \} $ when $ b\to0 $.
\begin{widetext}
\begin{gather}
1/\mathrm{Inf}\{ \lambda_\perp \} = \frac{4}{3} \sinh\alpha\sum_{n=1}^{\infty}\frac{n(n+1)}{(2n-1)(2n+3)} \left[ \frac{2\sinh(2n+1)\alpha + (2n+1)\sinh2\alpha}{4\sinh^2(n+1/2)\alpha - (2n+1)^2\sinh^2\alpha}-1 \right],\label{eq:inf} \\
1/\mathrm{Sup}\{ \lambda_\perp \} = \frac{4}{3} \sinh\alpha\sum_{n=1}^{\infty}\frac{n(n+1)}{(2n-1)(2n+3)} \left[ \frac{4\cosh^2(n+1/2)\alpha + (2n+1)^2\sinh^2\alpha}{2\sinh(2n+1)\alpha - (2n+1)\sinh2\alpha}-1 \right],\\
\mathrm{Inf}\{ \lambda_\parallel \} = 1-\frac{9}{16}\tau_\perp+\frac{1}{8}\tau_\perp^3-\frac{45}{256}\tau_\perp^4-\frac{1}{16}\tau_\perp^5-\frac{83}{256}\tau_\perp^6, \\
\mathrm{Sup}\{ \lambda_\parallel \} =1-\frac{9}{16}\tau_\parallel+\frac{1}{8}\tau_\parallel^3-\frac{45}{256}\tau_\parallel^4-\frac{1}{16}\tau_\parallel^5-\frac{83}{256}\tau_\parallel^6,\label{eq:sup}
\end{gather}
\end{widetext}

\begin{figure}[htb]
\scalebox{0.5}{\includegraphics{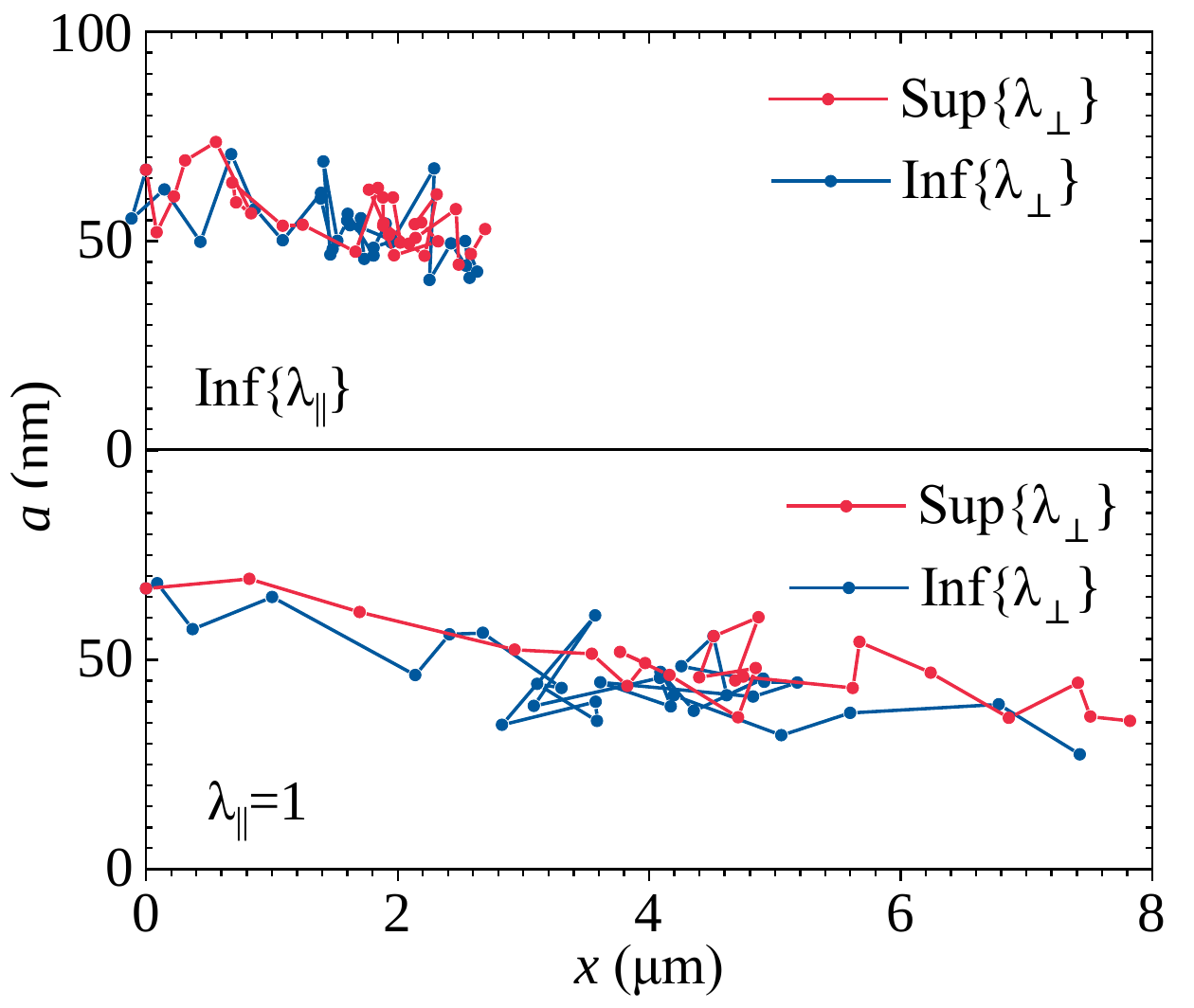}}
\caption{\label{fig:limitVelocity} (color online) The simulated sphere trajectory within 30 s, based on different mobility profiles. Time step $ \delta t=1\,\mathrm{ms} $ in calculations and 1 s for presented data. }
\end{figure}

In our considered case in the main text, where there is an equilibrium height $ a>0 $ ($ \tau_\perp<1 $), different $ \lambda_\perp $ in the available range makes negligible difference to the transport effect, but $ \lambda_\parallel $ significantly influences the transport velocity, as shown in \fref{fig:limitVelocity}. $ \mathrm{Inf}\{ \lambda_\parallel \}\ne0 $ when $ \tau_\perp<1 $ guarantees the robustness of the Casimir transport against variations in the hydrodynamic mobility profiles, and \fref{fig:limitVelocity} further confirms that even $ \mathrm{Inf}\{ \lambda_\parallel \} $ leads to observable transport behavior.

\bibliography{Casimir,NanoDynamics,meta,TO}